\documentclass{article}
\usepackage{spconf,amsmath,graphicx}
\usepackage{multirow,booktabs}
\usepackage{hhline}
\usepackage{mathtools}
\usepackage{amsmath,amssymb,amsfonts}
\usepackage{algorithmic}
\usepackage{comment}
\usepackage{subfig}
\usepackage{caption}
\usepackage{graphicx}

\title{Superresolution and Segmentation of OCT scans using  Multi-Stage adversarial Guided Attention Training}

\name{{\normalsize Paria Jeihouni, Omid Dehzangi, Annahita Amireskandari, Ali Dabouei, Ali Rezai, Nasser M. Nasrabadi}}

\address{{\footnotesize Rockefeller Neuroscience Institute, Computer Science \& Electrical Engineering, Ophthalmology \& Visual Sciences, West Virginia University, , USA}\\\textit{{\small \{\{pj00001, ad0046\}@mix, \{omid.dehzangi@, annahita.amireskandari@,  ali.rezai@\}hsc,
nasser.nasrabadi@mail\}.wvu.edu}}} 
\begin{document}
%
\maketitle
\begin{abstract}
Optical coherence tomography (OCT) is one of the non-invasive and easy-to-acquire biomarkers (the thickness of the retinal layers, which is detectable within OCT scans) being investigated to diagnose Alzheimer's disease (AD). This work aims to segment the OCT images automatically; however, it is a challenging task due to various issues such as the speckle noise, small target region, and unfavorable imaging conditions. In our previous work, we have proposed the multi-stage \& multi-discriminatory generative adversarial network (MultiSDGAN) \cite{jeihouni2021multisdgan} to translate OCT scans in high-resolution segmentation labels. In this investigation, we aim to evaluate and compare various combinations of channel and spatial attention to the MultiSDGAN architecture to extract more powerful feature maps by capturing rich contextual relationships to improve segmentation performance. Moreover, we developed and evaluated a guided mutli-stage attention framework where we incorporated a guided attention mechanism by forcing an L-1 loss between a specifically designed binary mask and the generated attention maps. Our ablation  study results on the WVU-OCT data-set in five-fold cross-validation (5-CV) suggest that the proposed MultiSDGAN with a serial attention module provides the most competitive performance, and guiding the spatial attention feature maps by binary masks further improves the performance in our proposed network. Comparing the baseline model with adding the guided-attention, our results demonstrated relative improvements of 21.44\% and 19.45\% on the Dice coefficient and SSIM, respectively. 
\end{abstract}
%

%
\section{Introduction}
\label{sec:intro}

AD is a progressive neurodegenerative disease that gradually declines memory and cognitive function. Previous studies have reported that the retina shares similar anatomical and physiological features with the brain, so it can be used as a possible biomarker for AD diagnosis in clinical practice \cite{sanchez2020evaluation}. Unlike current standard methods that are invasive and expensive for AD detection \cite{ferreira2011neuroimaging}, the thickness of the retina layer can be noninvasively assessed using high-resolution
images obtained with optical coherence tomography (OCT). Because of noise and  artifacts (e.g., eye motions, the vessel projection shadow), manual
segmentation of OCT images is a challenging task. Hence, it is imperative to program a method of OCT-based automatic retina layer segmentation.\\
\indent Convolutional neural networks (CNNs) have achieved state-of-the-art performance in a breadth of image segmentation tasks, and they have robust and nonlinear feature extraction capabilities \cite{sinha2020multi,9506495}. Although U-Net \cite{siddique2020u} is a common network for medical image segmentation, it has issues dealing with class imbalance labels. The main problem is the usage of cross-entropy (CE) loss \cite{murugesan2020context}. Since the foreground to background ratio is low in the medical images, using CE will learn a decision boundary biased towards the majority class, which would result in inaccurate segmentation. On the other hand, Generative Adversarial Networks (GANs) have been extensively used for various challenging medical segmentation tasks \cite{lei2020skin}. GANs have been quite prominent in learning deep representations and modeling high-dimensional data. Conditional GANs depict good performance translating data from one domain to another \cite{isola2017image}, \cite{zhu2017unpaired}, thus it is appropriate for semantic segmentation. In our previous works, we proposed a GAN-based domain translation and superresolution architecture that learns to increase the medical image resolution from low to high and learn to segment the retinal layers at the same time \cite{9506291,9412818, jeihouni2021multisdgan}. This particular type of GAN considers multiple stages of output from different layers of the network. Each intermediary output from the multi-stage is subjected to various discriminators \cite{jeihouni2021multisdgan}.
\indent Attention modules were widely used to boost segmentation performance \cite{liu2020covariance}. The attention module allows the network to focus on the most relevant features without additional supervision, avoiding using multiple similar feature maps and highlighting salient features that are useful for a given task. Channel attention selects meaningful features at channel dimension, and spatial attention calculates the feature representation in each position by the weighted sum of the features from all the other positions \cite{sagar2021dmsanet}. Previous studies depict the importance of attention modules on improving the performance of OCT image segmentation \cite{li2020pyramid}; however, to the best of our knowledge, it is the first time that a GAN-based attention module has been employed for OCT segmentation. \\
\indent In this paper, we aim to investigate the impact of incorporating attention mechanism in the MultiSDGAN framework \cite{jeihouni2021multisdgan}. We aim to evaluate and compare various combinations of channel and spatial attention to capture rich contextual relationships to  extract more powerful feature maps and improve segmentation outcome. Therefore, we design a method to train the network based on multi-stage attention modules and assess if there is any improve in segmentation performance. Moreover, this study take a step further and add regularization on top of the spatial attention feature maps to focus the attention on our region of interest, which we refer to that as the guided attention module. In this way, the guided attention modules are added to the generator by forcing the L-1 loss between a specifically designed binary mask and the attention maps generated at different layers of the network and we investigate its effectiveness in improving the final results. To the best of our knowledge, it is the first time that this study has been investigated. The rest of this paper is focused on introducing and evaluating the impact of various combinations of channel and spatial attention in multi-stages with self and guided attention approach on the MultiSDGAN model.
\section{Data acquisition and preprocessing}
Participants are recruited based on referrals to current patients at the memory disorders clinic or geriatric clinic at the  West  Virginia  University (WVU). An ophthalmologist conducted a complete eye exam on all the subjects, including visual acuity, intraocular pressure, pupillary reaction, and dilated fundus exam. The Heidelberg Spectralis OCT (Heidelberg Engineering Inc., Heidelberg, Germany) was used to obtain the OCT of the macula and the optic nerve head. 

Data collection was initiated with normal aging patients (age: 55+). The Ophthalmology Department at the WVU medicine provided the OCT images of 55 subjects, each having 19 scans and six subjects had one extra OCT. In total, our data-set has 1,045 images. These are 2-D scans, each group of 19 constitute one 3-D scan of the macula. Each image was meticulously labeled each image for the 7 innermost layers by an expert in the field. Finally, all patient data was de-identified prior to analysis based on  the WVU Institutional Review Board (IRB)  approved under the study ID: 1910761036. Horizontal flipping, spatial translation, and rotation are among the methods we employed to augment the data-set\cite{roy2017relaynet}. Also, to increase the data-set size synthetically, we used a moving crop window approach of size 224x224, which was moves on the image  with 75\% overlap.

\section{Methodology}
\subsection{MultiSDGAN}
\begin{figure}[t]
\centering
\includegraphics[width=9.6cm]{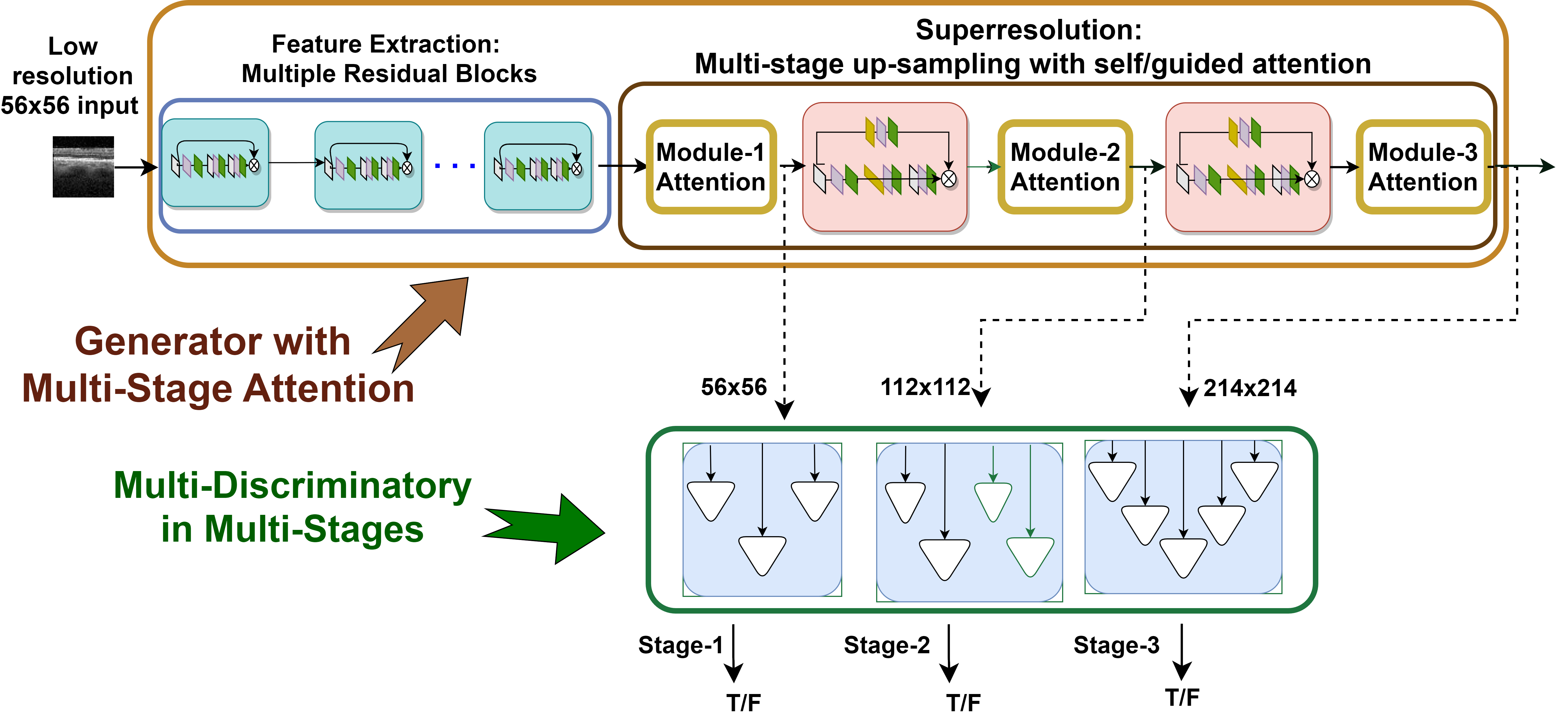}
\caption{{\small Our Multi-Stage Multi-Discriminatory GAN (MultiSDGAN) architecture is used as the framework for superresolution and segmentation. Multi-stage generator $G$ consists of several residual blocks and transposed blocks with added attention mechanisms in various stages of superresolution. The multi-discriminatory modules provide scrutiny at different patch levels.}}
\label{fig:main}
\end{figure}

Generative adversarial network contains of two subnetworks: the generator and the discriminator. Fig.\ref{fig:main} illustrates our GAN-based domain translation framework, MultiSDGAN. MultiSDGAN adopts and modifies ResNet as its generator architecture. 
The generator that is employed in this architecture has two major parts. The first part is being used for extracting features, and the second part superresolves the images to a certain scale. To achieve the superresolution, a transposed bottleneck block was designed to be added to the generator. One important feature of this generator is its multi-stage output, which is basically extracting outputs from different intermediatory layers of the network, rather than only the final layer as suggested in \cite{zhang2018stackgan++}.  Additionally, multiple discriminators are being used to enhance the discriminatory aspects of the GAN. Each of these discriminators is a PatchGAN \cite{isola2017image}, in which a convolutional neural network classifies an image as fake or real by focusing on penalizing it at the scale of local image patches of size $N \times N$.

\subsection{Attention module}
Attention mechanisms allow humans to selectively focus on key information while ignoring other irrelevant information. Through the attention module, deep CNN can extract more critical and discriminative features for the target task, and enhance the robustness of the network model \cite{tong2021ascu}. 
\subsubsection{Channel attention module}
\begin{figure}[t]
\centering
\includegraphics[width=8.5cm]{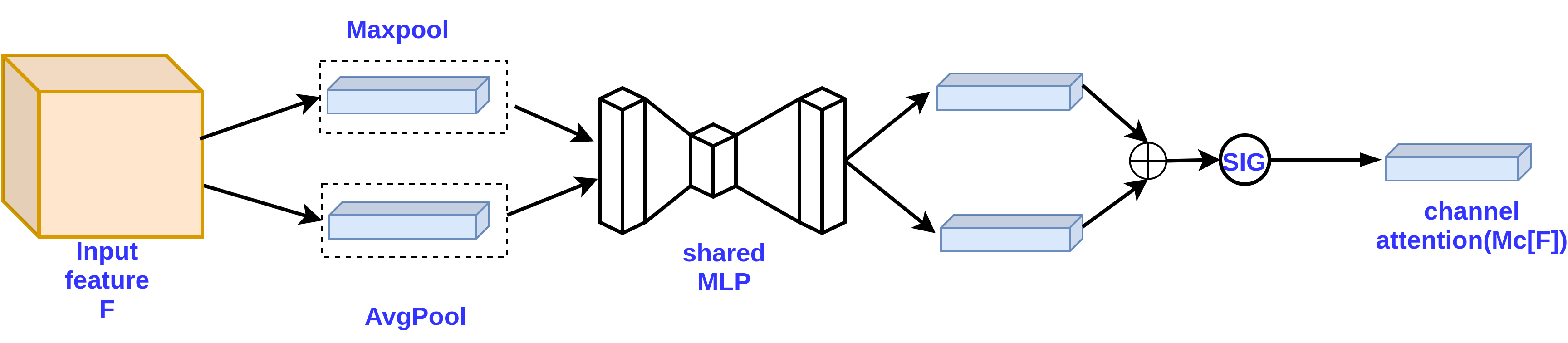}
\caption{Channel attention module.}
\label{fig:channel}
\end{figure}
The channel attention module is used to selectively weight the importance of each channel and thus produces best output features. This helps in reducing the number of parameters of the network.  To compute the channel attention, the spatial dimension of the input feature map was squeezed by average-pooling and Max-pooling \cite{woo2018cbam}. Fig. \ref{fig:channel} illustrates the channel attention mechansim. In short, the channel attention is computed as:

\begin{equation}
   M_{c}(F) = \sigma(MLP[AvgPool(\textup{F})]) +\sigma(MLP[MaxPool(\textup{F})]
\end{equation}

\subsubsection{{\small Spatial attention module}}
\begin{figure}[t]
\centering
\includegraphics[width=6.0cm]{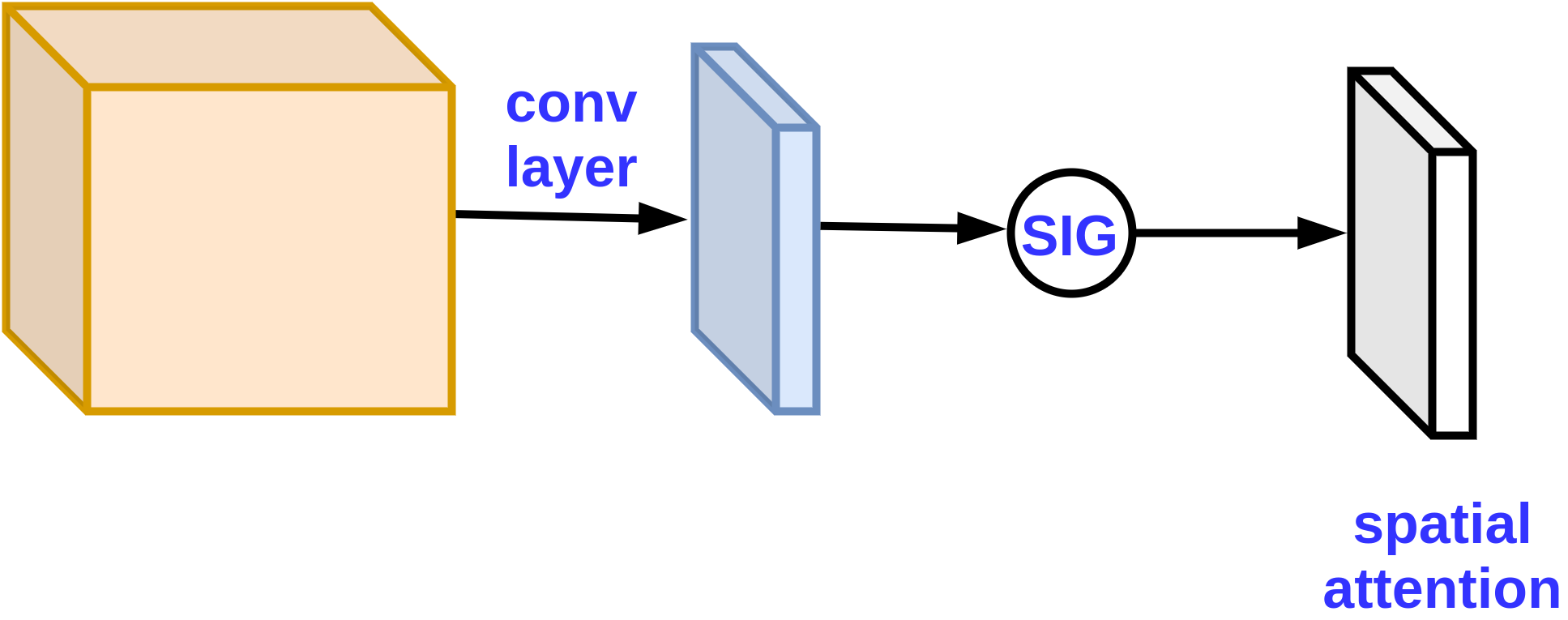}
\caption{{\small Spatial attention module.}}
\label{fig:spatial}
\end{figure}
This module is designed to learn the spatial dependencies in the feature maps. Specifically, a depth-wise convolution is used to extract information to have distant vision over the feature maps. To compute spatial attention, we apply  $1 \times 1$ convolution instead of max-pooling and avg-pooling to decrease the depth of the feature maps. In this way, the model learns how to shrink the depth by keeping the most relevant information. Fig. \ref{fig:spatial} illustrates the spatial attention mechanism. In short, the spatial attention is computed as:
\begin{equation}
    M_{s}(F) = \sigma(conv_{1x1}(\textup{F})).
\end{equation}

\begin{figure}[t]
\centering
\subfloat[\\Attention modules combined in serial.]{\includegraphics[width=0.45\textwidth]{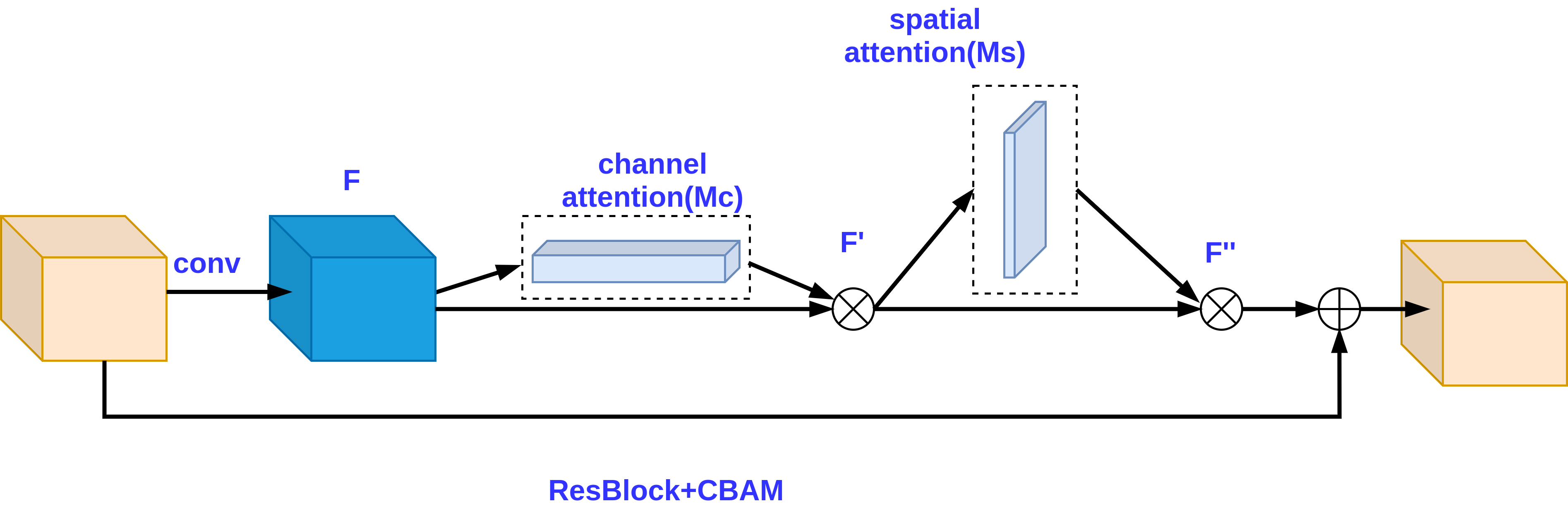}%
\label{fig:serial}}

\subfloat[\\Attention modules combined in parallel.]{\includegraphics[width=0.33\textwidth]{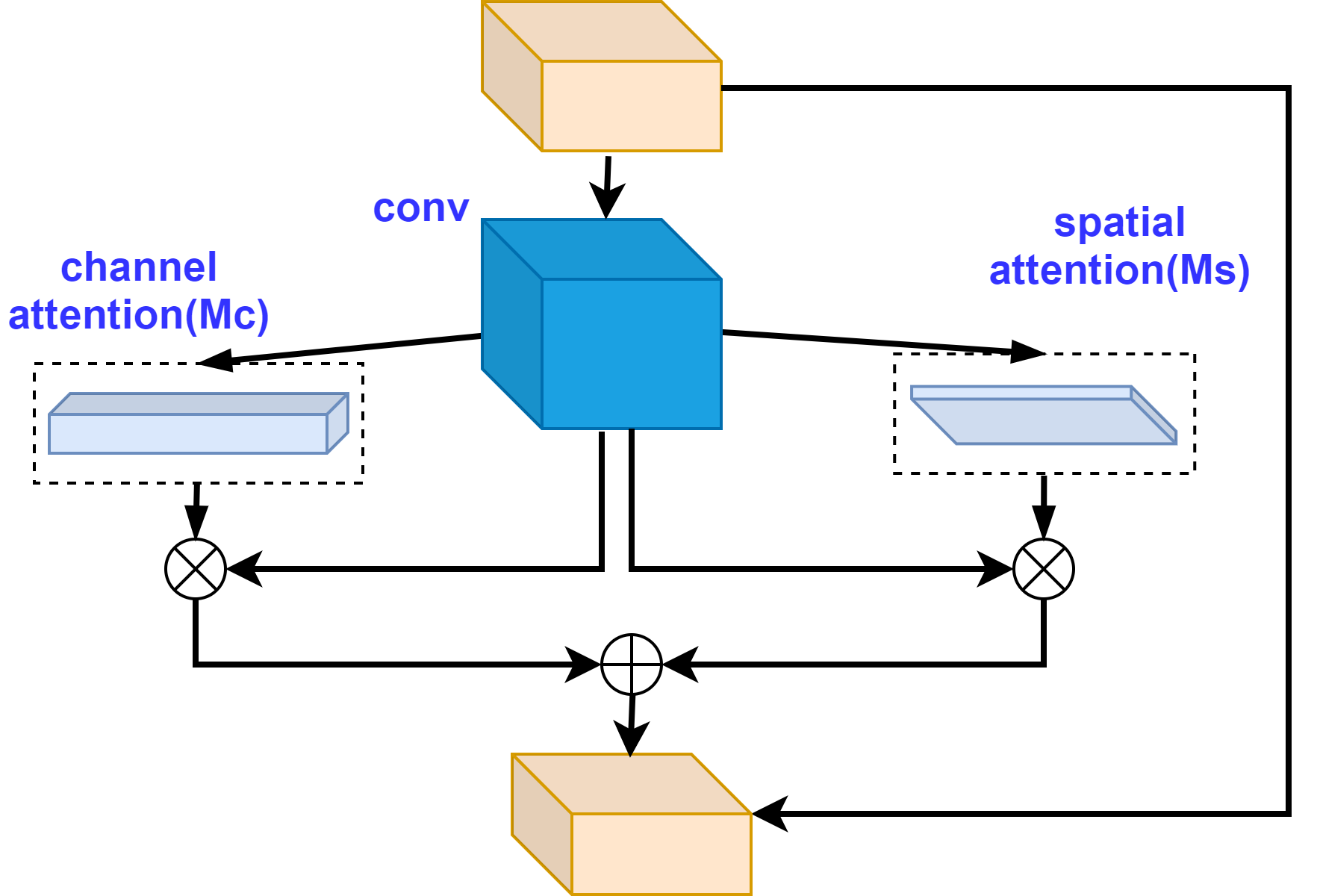}%
\label{fig:par}}

\caption{{\small Sequential and parallel combinations of channel and spatial attention modules.}}
\label{fig:combin}
\end{figure}

As we discussed previously, we want to investigate the impact of the serial and the parallel attention module beside individual channel and spatial attention. Fig. ~\ref{fig:serial} and ~\ref{fig:par} depicts the architecture of sequential and parallel attention. 

\begin{table*}[t]
\caption{SSIM, Dice coef and L-1 comparison among various combinations of attention modules.}
\label{table1}
{\renewcommand{\arraystretch}{1.5}
\centering
\resizebox{1.01\textwidth}{!}{%
\begin{tabular}{||c|c|c|c|c||c||c|c||c||c|c||c||}
\hline\hline
\multicolumn{3}{||c|}{Model}                                                                                                             & \multicolumn{3}{||c||}{Dice Coeff.}                                                                                                                                                                 & \multicolumn{3}{c||}{SSIM}                                                                                                                                                                 & \multicolumn{3}{c||}{L-1}                                                                                                                                                                  \\ \hline\hline
\multicolumn{3}{||c|}{Measure}                                                                                                           & \begin{tabular}[c]{@{}c@{}}Last Stage\\ Attention\end{tabular} & \begin{tabular}[c]{@{}c@{}}Multi-Stage\\ Attention\end{tabular} & \begin{tabular}[c]{@{}c@{}}No\\ Attention\end{tabular} & \begin{tabular}[c]{@{}c@{}}Last Stage\\ Attention\end{tabular} & \begin{tabular}[c]{@{}c@{}}Multi-Stage\\ Attention\end{tabular} & \begin{tabular}[c]{@{}c@{}}No\\ Attention\end{tabular} & \begin{tabular}[c]{@{}c@{}}Last Stage\\ Attention\end{tabular} & \begin{tabular}[c]{@{}c@{}}Multi-Stage\\ Attention\end{tabular} & \begin{tabular}[c]{@{}c@{}}No\\ Attention\end{tabular} \\ \hline
\multirow{4}{*}{\begin{tabular}[c]{@{}c@{}}MultiSDGAN\\ with attention\\ mechanism\end{tabular}} &
\multicolumn{2}{c|}{Channel}         &                    0.9076±0.007                                           &                    0.9102  ±0.008                                            &                 0.9016±0.004                                       &       0.9012±0.0048                                                         &      0.9086±0.0038                                                             &        0.8987±0.0051                                                &  0.020 ±0.0012                                                               &            0.019 ±0.0012                                                       &          0.021 ±0.0011                                              \\ \cline{2-12} 
                                                                                                 & \multicolumn{2}{c|}{Spatial}           &    0.9105 ±0.005                                                          &    0.9076 ±0.008                                                           &   0.9016±0.004                                                      &                                                0.9038±0.0022              &                     0.9093±0.0034                                              &         0.8987±0.0051                                                 &                 0.020±0.0011                                                &                        0.018±0.0012                                          &               0.021±0.0011                                          \\ \cline{2-12} 
                                                                                                 & \multirow{2}{*}{Ch\&Sp} & Parallel   &                0.9132±0.006                                                  &                            0.9134±0.002                                    &          0.9016±0.004                                               &                                    0.9087±0.0024                            &     0.9145±0.0036                                                            &       0.8987±0.0051                                                   &       0.018±0.0012                                                         &         0.017±0.0015                                                        &                  0.021±0.0011                                       \\ \cline{3-12} 
                                                                                                 &                         & Sequential &   0.9158±0.002                                                             &   \textbf{0.9187±0.0011}                                                               &         0.9016±0.004                                                &               0.9125±0.0034                                                 &             \textbf{0.9153±0.0024}                                                    &          0.8987±0.0051                                               &                  0.018±0.0011                                              &                 \textbf{0.016±0.0010}                                               &         0.021±0.0011                                                \\ \hline\hline
\end{tabular}
}
}
\end{table*}

\subsubsection{Guided Attention Module}
\begin{figure}[!]
\centering
\subfloat[\\label.]{\includegraphics[width=0.15\textwidth]{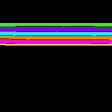}%
\label{fig:self}}
\hfil
\subfloat[\\Binary Guided attention mask.]{\includegraphics[width=0.15\textwidth]{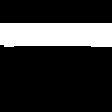}%
\label{fig:mask}}
\hfil
\caption{{\small The attention is guided to by focus on the region of interest shown in (a) by enforcing the attention mask in (b).}}
\label{fig:comparioson}
\end{figure}

As discussed earlier, positional information of images is the main focus of spatial attention, and it can detect the spatial relationship between the input features \cite{lee2020centermask}. To improve the performance of spatial attention even further, we adopt a binary mask to guide the feature maps for spotlighting the region of interest. Fig.\ref{fig:mask} depicts the binary mask in which only the region of interest (i.e., inner retina layers) is white. 
\subsection{Loss function}
Similar to the MultiSDGAN model \cite{jeihouni2021multisdgan}, during the training, the proposed network weights are updated based on the Dice, SSIM and L-1 losses that are obtained via the following formulas. y is our ground truth and G(\textup{x}) is the generator output.
\begin{equation}
   L_{L-1}(G) = \|\textup{y}-G(\textup{x})\|_{1},
\end{equation}
\begin{equation}
L_{Dice}(G) = 1-\frac{2\sum_{i=1}^{n}G(x_{i})y_{i}}{\sum_{i=1}^{n}G(x_{i})^2+\sum_{i=1}^{n}y_{i}^2},
    \label{equ:dice loss}
\end{equation}
\begin{equation} 
SSIM(G(\textup{x}),\textup{y}) = \frac{(2\mu_{G(\textup{x})}\mu_{\textup{y}}+c_{1})(2\sigma_{G(\textup{x}),\textup{y}}+c_{2})}{(\mu_{G(\textup{x})}^2+\mu_{\textup{y}}^2+c_{1})(\sigma_{G(\textup{x})}^2+\sigma_{\textup{y}}^2+c_{2})}.
\end{equation}

\section{Experiments \& Results}
In this section, the proposed architecture will be analyzed using various combinations of attention modules. Then, the whole architecture including applying attention modules on multiple stages will be discussed, and the results will be compared. In the end, the effect of using attention mask will be explained. The parameters set in the network are as follows:
Learning rate= 0.001, loss function= \textit{Dice, SSIM, L-1}, optimizer function= \textit{Adam}, batch size= 8, number of epochs= 200. Furthermore, we have divided the data-set into two parts, the train set and the validation set, where the percentage of the divisions are 80\% and 20\% (5-CV), respectively.

\subsection{Impact of self-attention mechanisms}  
The impact of four attention modules, namely, single spatial attention, single-channel attention, parallel attention, and Sequential attention, are investigated in this study. This comparison aims to find the best attention module for the ask in hand. In the first stage, we applied these attention modules to the last layer. As shown in Table \ref{table1}, the best-achieved performance belongs to the Sequential attention module consistently on the Dice coefficient, SSIM, and L-1 loss. The next best result is achieved by the parallel attention module. 

\subsection{Impact of multi-stage self-attention mechanisms} 
In the next stage, we applied the attention modules to all stages of suerresolution in the MultiSDGAN framework. As shown in Table \ref{table1}, the multi-stage extension (see Fig. 1) improves the evaluation criteria consistently. The best-achieved performance belongs to the sequential attention module in multi-stages and the parallel attention is the next best. 
\subsection{Impact of multi-stage guided-attention mechanisms} 
\begin{table}[t]
\caption{{\small SSIM, Dice coef and L-1 comparison among MultiSDGAN, MultiSDGAN with Sequential attention module and MultiSDGAN with Sequential guided attention module.}}
\label{table2}
\resizebox{0.5\textwidth}{!}{
{\renewcommand{\arraystretch}{2.3}
\begin{tabular}{|l|c|c|c|c|}
\hline
\multicolumn{2}{|c|}{\textbf{Model} }               & \textbf{Dice Coefficient} & \textbf{SSIM}& \textbf{L-1} \\ \hhline{|=|=|=|=|=|}
  \multicolumn{2}{|c|}{\textbf{RelayNet} }  &	0.8828±0.0017	& 0.8613±0.0023	& 0.027±0.0011 \\
\hhline{|=|=|=|=|=|}

\multirow{3}{*}{\parbox{1.9cm}{\textbf{MultiSDGAN
}}}  &     \textbf{No Attention} & 0.9016±0.003 & 0.8987±0.0051 & 0.021±0.0011 
\\
\hhline{|~|-|-|-|-|}
&  \textbf{Self-attention} & 0.9187±0.0011& 0.9153±0.0024& 0.016±0.0010 
\\
\hhline{|~|-|-|-|-|}
&  \textbf{Guided-attention} & \textbf{0.9227±0.0022}& \textbf{0.9184±0.0054}& \textbf{0.016±0.0009} 
\\
\hhline{|=|=|=|=|=|}
\end{tabular}
}
}
\end{table}
The reported results in Table \Ref{table1} depict that we get the best performance from the Sequential attention module. We further improved the results via guiding the training using a binary mask. As it can be seen in Table \Ref{table2}, applying binary mask further improved the performance of Sequential attention in comparison with the baseline without any attention mechanism. It demonstrated relative improvements of 21.44\% (p\_value\textless0.05, t-test on mean differences) and 19.45\% (p\_value\textless0.05) on the Dice coefficient and SSIM, respectively. Also, Table \Ref{table2} demonstrate that all variants of our proposed adversarial attention mechanisms provide improved results in comparison with RelayNet \cite{roy2017relaynet}, as a strong baseline model (with the impressive highest relative improvement of 41.16\% and p\_value\textless0.01).

\begin{figure}[t]
\centering
\subfloat[\\Trained self-attention feature map.]{\includegraphics[width=0.16\textwidth]{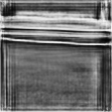}%
\label{fig:self}}
\hfil
\subfloat[\\Trained guided-attention feature map.]{\includegraphics[width=0.16\textwidth]{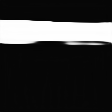}%
\label{fig:guided}}
\hfil
\caption{The final trained attention feature maps for the cases of (a) self- and (b) guided-attention.}
\label{fig:comparioson}
\end{figure}

\section{Conclusion}
In this paper, we proposed a new feature on our MultiSDGAN segmentation network that can refine features based on various attention modules. Experiments on our own data-set demonstrated the effectiveness of the attention mechanisms on MultiSDGAN. Sequential combination and guided attention mechanism provided the best empirical results by reducing the redundancy in model training. We aim to design attention modules effectively and capture more discriminative features for semantic inference as our future direction.
{\small
\bibliographystyle{IEEEtran}
\bibliography{strings}
}

\end{document}